\documentclass[12pt]{article}

\usepackage{scicite}

\usepackage{times}
\usepackage{graphicx}

\newenvironment{sciabstract}{%
\begin{quote} \bf}
{\end{quote}}

\newcounter{lastnote}
\newenvironment{scilastnote}{%
\setcounter{lastnote}{\value{enumiv}}%
\addtocounter{lastnote}{+1}%
\begin{list}%
{\arabic{lastnote}.}
{\setlength{\leftmargin}{.22in}}
{\setlength{\labelsep}{.5em}}}
{\end{list}}

\newcommand{\micron}{\mbox{$\mu{\rm m}$}}
\newcommand{\arcsec}{\mbox{$^{\prime \prime}$}}

\def\lesssim{\mathrel{\hbox{\rlap{\hbox{%
 \lower4pt\hbox{$\sim$}}}\hbox{$<$}}}}
\def\gtrsim{\mathrel{\hbox{\rlap{\hbox{%
 \lower4pt\hbox{$\sim$}}}\hbox{$>$}}}}
\def\sun{\hbox{$\odot$}}

\newcommand\mnras{MNRAS}
\newcommand\nat{Nature}%
\newcommand\aj{AJ}%
\newcommand\apj{ApJ}%
\newcommand\apjl{ApJ}%
\newcommand\aap{A\&A}%
\newcommand\pasp{PASP}%

\newcommand{\Msun}{\mbox{$M_{\sun}$}}
\newcommand{\Mearth}{\mbox{$M_{\oplus}$}}

\newcommand{\perone}{\mbox{$^{-1}$}}

\newcommand{\eg}{e.g.}
\newcommand{\ie}{i.e.}
\newcommand{\cf}{cf.}

\newcommand{\degs}{\mbox{$^{\circ}$}}

\newcommand{\bPic}{\hbox{$\beta$}~Pic}

\title{Substructure in the Circumstellar Disk\\
around the Young Star AU Mic (GJ 803)}

\author{Michael C. Liu\\
\normalsize{Institute for Astronomy, University of Hawaii,}\\
\normalsize{2680 Woodlawn Drive, Honolulu, HI 96822, USA}\\
\normalsize{E-mail:
  {\tt mliu@ifa.hawaii.edu}}\\
\\
\normalsize{Accepted for publication in {\it Science}}\\
\normalsize{August 4, 2004}\\
}

\date{}

\begin{document}

\parskip 1ex

\maketitle

\begin{sciabstract}
Keck adaptive optics imaging with a physical resolution of 0.4~AU
resolves the inner (15--80~AU) disk of AU~Mic (GJ~803), the nearest
known scattered light disk to Earth.  The inner disk is asymmetric and
possesses a sharp change in structure at 35~AU.  The disk also shows
spatially localized enhancements and deficits at 25--40~AU separations.
The overall morphology points to the influence of unseen larger bodies
and resembles structures expected from recent planet formation.  AU~Mic
is coeval with the archetypical debris disk system \bPic, and the
similarities between their two disks point to synchronous disk
evolution.  Multiple indications of substructure appear to be common in
circumstellar disks at an age of $\approx$12~Myr.
\end{sciabstract}

\section{Introduction}

After dissipation of their primordial disks of gas and dust, many stars
develop debris disks, which are composed solely of collisionally
regenerated dust.  Debris disks have been identified by their thermal
emission at IR and sub-millimeter wavelengths
\cite{1993prpl.conf.1253B,2000prpl.conf..639G}.  Their spectral energy
distributions (SEDs) convey only a limited information about the extant
physical processes.  In this regard, the large resolved disk around
\bPic\ has been a goldmine for scrutinizing the structure, composition,
and dynamics of a debris disk (\eg, \cite{1984Sci...226.1421S};
\cite{1997AREPS..25..175A} and references therein;
\cite{1997MNRAS.292..896M,2000ApJ...530L.133K,2003ApJ...584L..33W,2003ApJ...584L..27W,2004A&A...413..681B}).
However, a broader understanding has been hampered by the small number
of systems that are spatially resolved.

The newly discovered disk around the young star AU~Mic (GJ~803) offers a
promising opportunity to examine the debris disk phenomenon.  This
well-studied flare star is among the youngest known M~dwarfs in the
solar neighborhood, with an estimated age of 12$^{+8}_{-4}$~Myr
\cite{1999ApJ...520L.123B,2001ApJ...562L..87Z} and a distance of only
$9.94\pm0.13$~pc from Earth \cite{1997A&A...323L..49P}.  A recent
search of nearby young ($\sim$10--50~Myr) stars identified AU~Mic as a
bright sub-mm source, possessing 0.01~\Mearth\ of cold (40~K) dust with
a fractional luminosity of $L_{dust}/L_{star} = 6\times10^{-4}$
\cite{2004astro.ph..3131L}.  Follow-up $R$-band (0.65~\micron)
coronagraphic imaging discovered that AU~Mic has a large disk seen in
scattered light \cite{klm03}, the closest such disk to Earth.
The seeing-limited discovery images detected the disk as close to the
star as the outer edge of the focal plane mask, which was 5\arcsec\
(50~AU) in radius.  Little is known about AU~Mic's disk inside of 50~AU,
a scale which corresponds to edge of the classical Kuiper belt in our
own solar system \cite{2000prpl.conf.1201J}.

\section{Observations}

AU~Mic was observed on 27--28~June~2004 UT using the adaptive optics
(AO) system on the Keck~II 10-m~telescope \cite{2000PASP..112..315W}
with the facility coronagraphic camera NIRC2 and the $H$-band
(1.63~\micron) filter.  Conditions were clear, and the AO-corrected
images have a full width at half maximum (FWHM) of 0.04\arcsec\
(0.4~AU).  Photometric calibrations were based on IR standard stars
\cite{1998AJ....116.2475P} and 2MASS photometry of the star 21\arcsec\
to the southwest of AU~Mic.  The edge-on disk of AU~Mic is remarkably
bright, noticeable in individual raw images.  No similar features were
seen in images of other stars obtained as a control sample.

A "roll subtraction" technique was developed to remove the point spread
function (PSF) from the images, thereby enabling study of the inner disk
against the bright glare of the central star \cite{science-SOM}.  This
technique is similar to that used for observing programs with the {\sl
  Hubble Space Telescope}
\cite{2000ApJ...539..435H,2003SPIE.4860....1S}.  The AU~Mic disk is
detected in the Keck AO imaging from 15--80~AU in projected separation
(Fig.~1).  The disk midplane is resolved, with an observed FWHM of
2.0--2.5~AU inside of 50~AU.  The innermost ($<$15~AU) regions
are dominated by PSF subtraction residuals and are inaccessible to study
in this dataset.

\section{Properties of the Inner Disk}

The AU~Mic disk shows two large-scale asymmetries.  (1) The midplanes
are different sizes.  The outer isophotes of the SE side are
$\approx$10\% smaller than those of the NW side.  (2) The disk midplanes
are not aligned.  From 2--5\arcsec\ in radius, their position angles
(PA) are 129.3$\pm$0.8\degs\ and 311.4$\pm$1.0\degs, as measured east of
north.
The relative tilt from the $H$-band data is 2.1$\pm$1.3\degs, consistent
with the 6$\pm$3\degs\ tilt seen at $R$-band for the outer disk
\cite{klm03}.  Unlike the other asymmetries discussed in this paper,
this tilt asymmetry can be explained by the intrinsic dust scattering
properties \cite{1995AJ....110..794K}, rather than a structural
asymmetry in the disk.

The disk midplanes do not follow a constant position angle.  Both
midplanes curve slightly to the north, with a $\approx$1\degs\ change in
orientation seen in the $H$-band imaging.  This bowing is suggestive of
an interaction between the disk and the local interstellar medium
\cite{1997ApJ...490..863A}.  However, the direction of AU~Mic's proper
motion is nearly aligned with the disk midplane, which is not consistent
with this idea.  The bowing may reflect the internal dynamics of the
disk.

The radial surface brightness profile $f_\nu$ is well-described by a
power law, $f_\nu \propto r^{\alpha}$, where $r$ is the projected
separation (Fig.~2).  From 35--60~AU separation, $\alpha = -4.4\pm0.4$
and $-4.4\pm0.3$ for the SE and NW sides, respectively.  These are
somewhat steeper than the $R$-band measurements from 50--210~AU, which
have slopes of $-$3.6 to $-$3.9 \cite{klm03}, though the different
angular resolution of the two datasets impedes direct comparison
(0.04\arcsec\ for $H$-band, 1.1\arcsec\ for $R$-band).  In the innermost
disk, there is a change in slope at 35~AU, with $\alpha = -1.0\pm0.3$
and $-1.4\pm0.3$ for the SE and NW sides, respectively, as measured from
20--35~AU separation.  Also, there is an indication that the break
occurs at slightly different radii for the two midplanes, $\approx$32~AU
for the SE side and $\approx$38~AU for the NW side, pointing to a
nonaxisymmetric structure.

In addition to the large-scale asymmetries, the Keck AO imaging reveals
smaller-scale asymmetries in the disk at 25--40~AU, both radially and
vertically.  These features are spatially resolved, being broader than
the PSF; they lie outside the region of significant PSF subtraction
residuals; and they are consistent in independent subsets of the data.
Because the disk is seen nearly edge-on, the true physical prominence of
the substructure is diminished by the smooth component of the disk along
the line of sight.

The most obvious substructure resides in the SE midplane (Figs.~1
and~3).  The SE side contains at least two radial enhancements, one at
25~AU and another at 31~AU.  There is also a relative deficit in the
scattered light at 29~AU.  The NW side shows an enhancement at 25~AU
aligned with the SE feature at 25~AU, indicative of a limb-brightened
ring of material.  However, no other obvious counterparts for the SE
features are seen in the NW side, indicating structures with non-zero
eccentricity and/or incomplete aziumuthal extent (\eg, clumps).  Such
nonaxisymmetric structures are most naturally explained by the dynamical
influence of unseen planets
\cite{1999AJ....118..580L,2000ApJ...537L.147O}.

The disk also possesses vertical substructure (Fig.~4).  The prominent
SE lumps at 25 and 31~AU reside at different elevations.  Furthermore,
the NW side shows a local enhancement at 37~AU which is displaced from
the inner midplane.  The micron-sized dust grains responsible for the
scattered light are removed by collisions and/or Poynting-Robertson drag
on timescales shorter than the age of the star \cite{klm03}.
Therefore, the observed vertical substructure may originate from the
inclined ($\gtrsim$1\degs) orbits of larger unseen bodies, either the
parent bodies which collisionally produce the dust or planets which
gravitationally perturb the dust.

\section{Comparing Disk Structure: AU~Mic vs \bPic}

The debris disk archtype \bPic\ is the best-studied spatially resolved
disk system.  Many of the structural features present in the \bPic\ disk
are also found in the AU~Mic disk:

(1) Both disks have unequal-sized midplanes.  In the case of \bPic, the
NE side is larger than the SW side
\cite{1993ApJ...411L..41G,1995AJ....110..794K}, an effect which has
been attributed to eccentricity perturbations driven by a substellar
companion \cite{1988A&A...203L..13W}.

(2) The surface brightness profiles for the \bPic\ and AU~Mic disks are
similar, with steep outer profiles ($\alpha\approx -4\ \rm{to}\ -5$) and a
strong flattening ($\Delta\alpha\approx2-3$) in the inner profile.  For
\bPic, the profile changes markedly inside of 100~AU compared to the
outer disk
\cite{1993ApJ...411L..41G,1995AJ....110..794K,2000ApJ...539..435H}.  For
AU~Mic, the change occurs at 35~AU.

The steep slope of the \bPic\ outer disk has been modeled as dust
originating from an inner collisional planetesimal disk and then
radiatively driven outwards (\cite{2001A&A...370..447A}; \cf,
\cite{1996A&A...307..542L}).  
The strong flattening at 100~AU demarcates the outer extent of the
\bPic\ planetesimal disk.  By analogy, the similar flattening of the
AU~Mic profile suggests a planetesimal disk of 35~AU extent.  This value
is consistent with the $\approx$17~AU inner radius inferred from the
IR/sub-mm SED \cite{2004astro.ph..3131L}.  And taken together, the data
suggest that AU~Mic's planetesimal disk is restricted to
$\approx$17--35~AU in radius.

The factor of~3 difference in the inferred sizes of the underlying
planetesimal disks can be understood in the context of different
agglomeration rates.  The time scale for planetesimal growth scales as
$t \propto P / \Sigma$, where $P$ is the orbital period and $\Sigma$ is
the surface density \cite{1987Icar...69..249L}.  For a disk profile of
$\Sigma \propto \Sigma_0 a^{-3/2}$, the growth time scale is then $t
\propto a^3 / (\Sigma_0 M_\star^{1/2})$, \ie, strongly dependent on the
orbital radius~$a$.  The two stars are coeval, with a factor of~4
difference in stellar mass.  Assuming the difference in the total disk
masses is reflected in the factor of~10 difference in the sub-mm
emitting dust masses \cite{2003astro.ph.11593S,2004astro.ph..3131L},
planetesimal growth should have proceeded to $\sim$2.7$\times$ larger
radii for \bPic\ compared to AU~Mic, in accord with the observational
estimates.

(3) The \bPic\ disk exhibits small-scale structures in its inner disk
which are radially confined and vertically displaced
\cite{2003ApJ...584L..27W}, similar to the AU~Mic disk.  These features
are naturally explained by radially localized structures in the dust
with non-zero eccentricities and inclinations (\eg, rings, clumps, and
gaps), perhaps arising from resonant interactions of multiple planets
\cite{2003ApJ...597..566T}.  The \bPic\ disk also displays a prominent
inner warp \cite{1997MNRAS.292..896M,2000ApJ...539..435H}, with a
3--4\degs\ tilt relative to the outer disk.  Such a strong warp is not
seen in the AU~Mic disk, though its detectability would depend on its
radial extent and orientation to the line of sight.

\section{Evolutionary Context}

Since its discovery, the singular nature of the large scattered light
disk around \bPic\ has raised the question of whether the system
represents a typical phase in early disk evolution
\cite{1997AREPS..25..175A,1998ApJ...505..897J} or an anomalous
occurence, \eg, caused by the gravitational perturbation of a passing
star \cite{2000ApJ...530L.133K}.  The discovery and characterization of
the coeval AU~Mic disk demonstrate that a common phase in disk evolution
involves optically thin, asymmetric, bright scattered-light disks with
multiple indications of substructure.  Given that AU~Mic and \bPic\ are
members of the same moving group
\cite{1999ApJ...520L.123B,2001ApJ...562L..87Z}, the high degree of
similarity between these two disks suggests synchronous evolution has
occurred. 

Dust with sufficient optical depth to produce detectable scattered light
spans a large range in radius around \bPic\ and AU~Mic, from as close as
$\approx$15~AU out to hundreds of AU.
This is in contrast to older ($\gtrsim$200~Myr) debris disk systems,
where the dust is confined to ring-like structures detected in sub-mm
thermal emission \cite{1998ApJ...506L.133G,1998Natur.392..788H} but not
in scattered light \cite{1992A&A...261..499S,1996AJ....111.1347K}.
Recent simulations of evolving planetesimal disks are in accord with
this observed morphological transformation from young dusty disks to old
dusty rings \cite{2004AJ....127..513K}.  However, the young
($\sim$8~Myr) A~star HR~4796A has its scattered light confined to a
single bright ring \cite{1999ApJ...513L.127S}, as opposed to a large
disk, so stellar age cannot be the only factor governing disk
morphology.

The spatially localized enhancements and deficits found in the AU~Mic
disk resemble the expected signposts of recent and/or ongoing planet
formation in young disks.  Simulations of planet formation by
agglomeration find that bright rings of dust arise from gravitational
stirring of planetesimals by recently formed planets of $\gtrsim$1000~km
radius \cite{2002ApJ...577L..35K,2004AJ....127..513K}.  Dark gaps occur
where the dust has been dynamically removed by planets or represent
regions shadowed by interior rings which are optically thick.  In this
interpretation, the multiple structures present in the AU~Mic disk argue
that planets massive enough to induce significant gravitational stirring
form contemporaneously over a range of radii.

Finally, the stellar masses of \bPic\ (2~\Msun) and AU~Mic (0.5~\Msun)
straddle that of solar-mass stars.  Hence, scrutiny of these two
well-resolved disks may provide a window into the early solar system.
The young Kuiper belt was about a factor of~100 more massive than its
current state \cite{1997AJ....114..841S}; its fractional dust
luminosity would have been around $10^{-3}$ to $10^{-5}$
\cite{2000prpl.conf.1201J,2003ApJ...598..626D}, comparable to the
\bPic\ and AU~Mic disks.
This Keck AO study reveals that multiple dynamical substructures are
common to optically thin disks at ages of $\approx$12~Myr.  These
structures may also reflect the dynamics that were active in the young
Kuiper belt.




\begin{scilastnote}
\item It is a pleasure to thank Eugene Chiang, Greg Herczeg, Mike Jura,
  Paul Kalas, John Krist, Jeff Linsky, and Bruce Macintosh for
  enlightening discussions.  We are very grateful to Antonin Bouchez,
  David LeMignant, Randy Campbell, Peter Wizinowich, and the staff of
  Keck Observatory for their assistance with the observations. This
  research has made use of the NASA/IPAC, 2MASS, and SIMBAD
  databases. We acknowledge support from NSF grant AST04-07441 and NASA
  grant HST-GO-09845.01-A.
\end{scilastnote}

\clearpage

\begin{figure}
\vskip -2in
\hskip -0.25in
\centerline{\includegraphics[width=5in,angle=90]{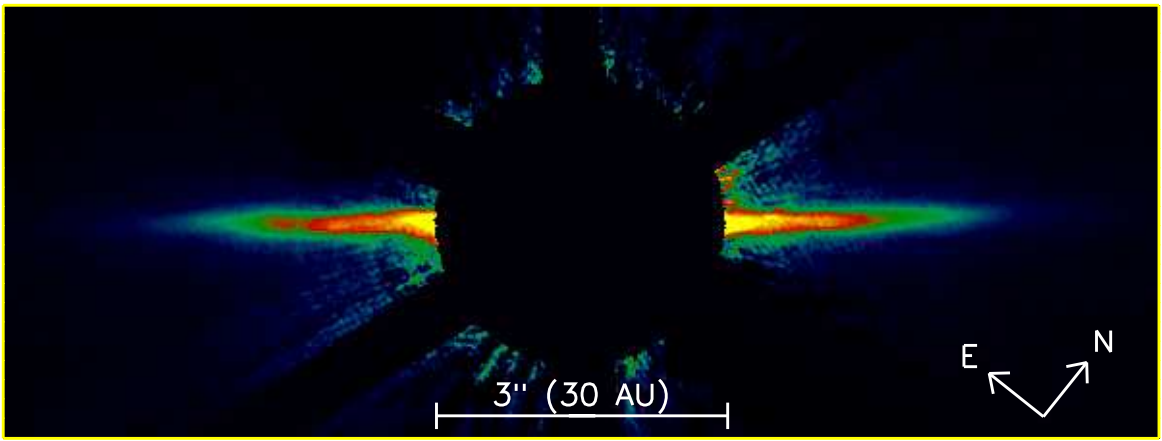}}
\vskip -5ex
\caption{\normalsize 
$H$-band (1.63~\micron) image of the AU~Mic dust disk obtained with the
Keck~II telescope. The image is 12\arcsec\ (119~AU) wide and 4.5\arcsec\
(45~AU) high. A software mask has been applied to block the PSF
subtraction residuals around the six diffraction spikes and the central
1.5\arcsec\ (15~AU) radius region.
\label{fig:plot-pretty}}
\end{figure}

\begin{figure}
\vskip -1in
\centerline{\includegraphics[width=4.5in,angle=90]{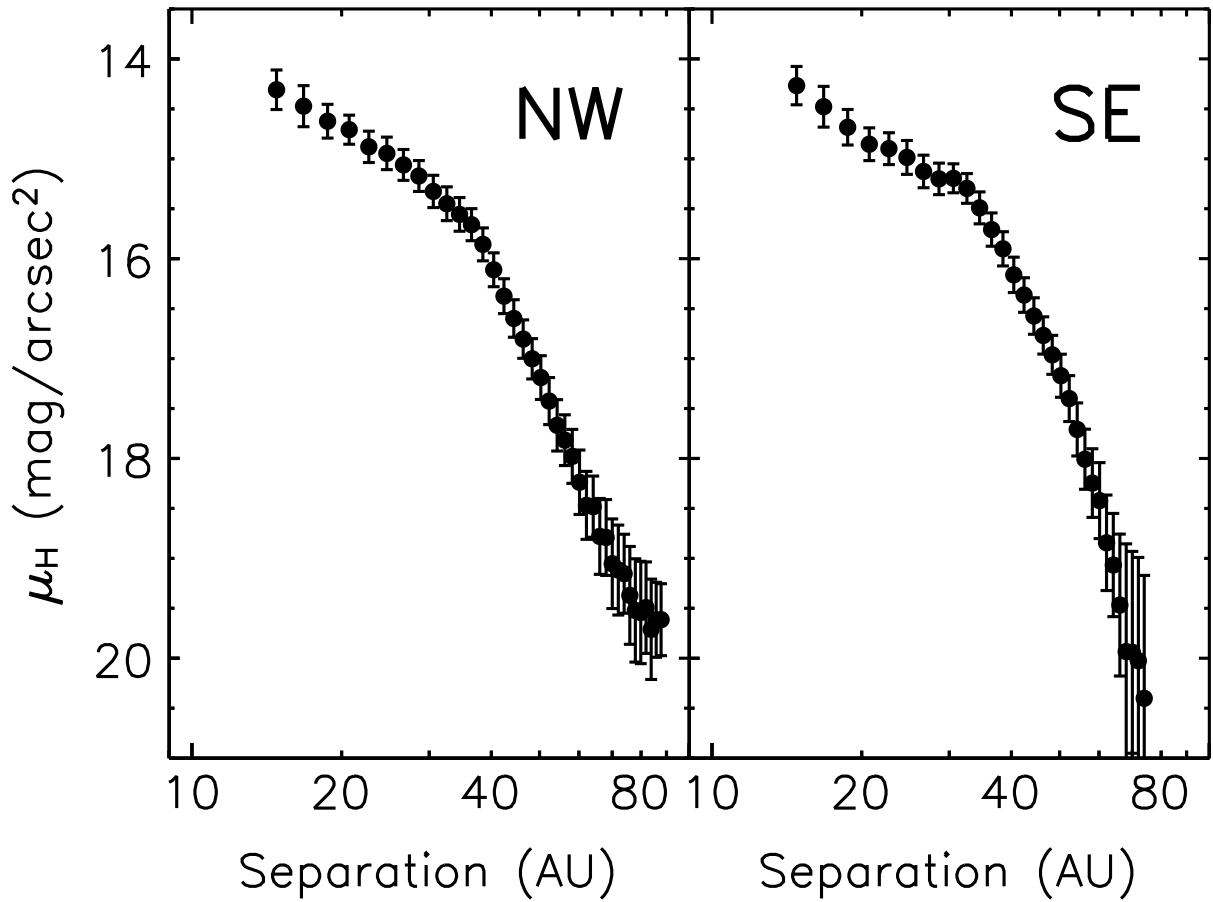}}
\vskip 4ex
\caption{\normalsize 
$H$-band (1.63~\micron) surface brightness
profile of the AU~Mic disk midplane derived from a photometry aperture
0.6\arcsec\ wide in the direction perpendicular to the midplane.
The NW side of the disk is larger than the SE side.  A break in the
surface brightness profile is seen at 35~AU.
\label{fig:plot-phot}}
\end{figure}

\begin{figure}
\vskip -1in
\centerline{\includegraphics[width=5in,angle=90]{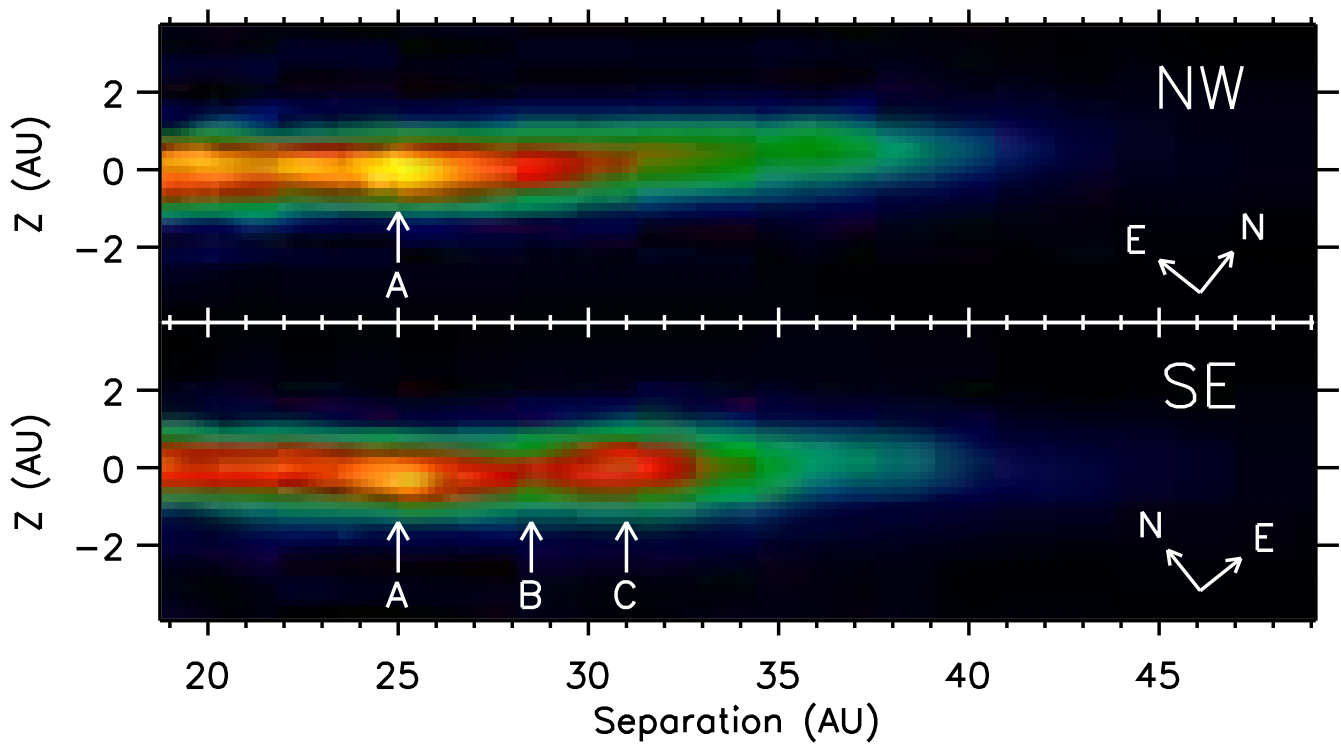}}
\vskip 2ex
\caption{\normalsize 
Radial substructure in the AU~Mic disk.  The SE
data have been mirrored about the axis perpendicular to the disk
midplane.  To highlight the substructure, each pixel has been multiplied
by its distance from the star, in order to compensate for the overall
decrease in disk flux with radius.  The data have been gaussian smoothed
to the image resolution of 0.04\arcsec\ (0.4~AU).  The data are oriented
with the SE midplane horizontal; the small relative tilt of the NW
midplane can be seen.
\label{fig:plot-lumps}}
\end{figure}

\begin{figure}
\vskip -1in
\centerline{\includegraphics[width=4.7in,angle=90]{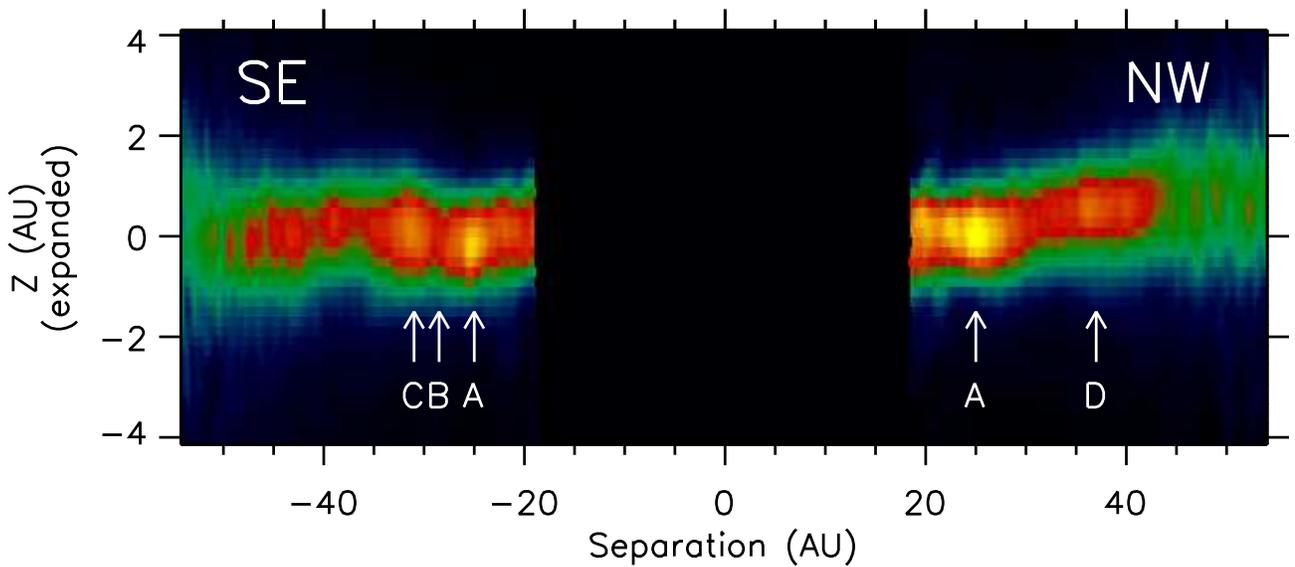}}
\vskip 2ex
\caption{\normalsize 
Vertical substructure in the AU~Mic disk.
  The plot's vertical axis has been expanded by a factor of 5.  The NW
  concentration at 25~AU is aligned with its SE counterpart, but the
  strong features at 25 and 31~AU in the SE midplane reside at different
  heights.  The elevated NW structure at 37~AU has no clear counterpart.
  To show the structure over a wide range of separations, the disk flux
  has been normalized by the radial surface brightness profile in Fig.~2
  (different than the normalization used for Fig.~3).  The data have
  been gaussian smoothed to the image resolution of 0.04\arcsec\
  (0.4~AU). The image orientation is the same as in
  Fig.~1.  
\label{fig:plot-warp}}
\end{figure}

\clearpage

\section*{SUPPORTING ONLINE MATERIAL:\\
Observing and Data Reduction Methods}

A combination of coronagraphy and a ``roll subtraction'' observing
technique were employed to maximize the dynamic range and contrast of
the observation, thereby enabling study of the scattered light disk
around AU~Mic against the bright glare of the central star.

The facility near-IR camera NIRC2 is equipped with a variety of
translucent circular masks in its first focal plane.  The 2\arcsec\
diameter mask was used for the AU~Mic observations.  The mask greatly
reduced the light from the PSF's central region, thereby avoiding
detector saturation and allowing for longer integrations.  The net
result was to increase the dynamic range of the instrument and the
observing efficiency.  Both the wide and narrow camera optics in NIRC2
were used, which provide pixel scales of
39.69$\pm$0.05~mas~pixel\perone\ and 9.94$\pm$0.05~mas~pixel\perone,
respectively.

A ``roll subtraction'' technique was developed to remove the PSF,
similar to methods used for observing programs with the {\sl Hubble
  Space Telescope}.  Taking advantage of the fact that Keck is an
alt-azimuth telescope, images were obtained such that the telescope
pupil orientation, and hence the Keck PSF, remained fixed relative to
the detector while the orientation of the sky rotated.  Thus, the
orientation of AU~Mic's disk changed on the detector as the star moved
across the sky.  Individual images were subtracted from each other to
remove the PSF, but leaving the disk.  The optimal scaling between two
images was determined by minimizing the residuals in an annular region
of high PSF flux, with the scaling factors being at most a few percent.

Individual roll-subtracted images were then aligned and rotated to a
common orientation; the diffraction spikes and the negative counterpart
images from the roll subtraction were masked; and a final mosaic was
constructed by median averaging.  One potential limitation of this
method is that the disk may overlap and thereby self-subtract if the
amount of rotation between the two images is small.  This effect was
avoided by observing AU Mic over a wide range of sky orientations,
resulting in 30--40\degs\ of rotation between subtracted pairs of
images.

Photometry errors are dominated by the residuals from roll subtraction.
To quantify these at each radius, we measured the scatter in photometry
from apertures placed at arbitrary position angles and using independent
subsets of the data.  These errors were added in quadrature to the
errors in the photometric calibration.  The resulting total errors are
plotted in Fig.~2.

\end{document}